\documentstyle[12pt,psfig,twoside,fleqn,qm99-wa98-flow]{article}

\newcommand{\AmS}{{\protect\the\textfont2
                    A\kern-.1667em\lower.5ex\hbox{M}\kern-.125emS}}
\title{Directed and Elliptic Flow in 158$A$GeV Pb+Pb Collisions}
\author{H. Schlagheck\address{Institut f\"ur Kernphysik, 
        Westf\"alische Wilhelms Universit\"at, \\ 
        Wilhelm Klemm Str. 9, 48149 M\"unster, Germany}
        for the WA98 collaboration}
\begin{document}
\pagestyle{empty}
\maketitle
\begin{abstract}
  Directed and elliptic flow of protons and positively charged
pions has been studied in the target fragmentation region using 
the Plastic Ball detector in the WA98 experiment. The results 
exhibit a strong dependence on centrality, rapidity, and transverse 
momentum.
  The rapidity dependence can be described by a Gaussian distribution.
The model comparisons reveal a large discrepancy of the flow strength
obtained from the data and the simulations. 
\end{abstract}

\section{Introduction}
  In high energy collisions it is expected that a high density 
interaction zone is formed. If this system thermalizes, the thermal
pressure will necessarily generate collective transverse 
expansion\cite{SIM-STOC-94-A,SIM-MATT-95-A}. 
If the initial state is azimuthally asymmetric, as
in semi-central collisions, this property may be reflected in the
azimuthal asymmetry of the final state particle distributions. 
The strength of collective flow will yield information on the 
nuclear equation of state during the expansion. Especially the
in-plane or out-of-plane character of the elliptic flow should
give important hints on the underlying mechanisms.

\section{Method}
  Two independent methods are used to determine the strength of
the collective flow. To be able to compare the result to previous
data the average transverse momentum  $\langle p_x \rangle$ method
is used. In this method the transverse momentum $p_T$ of each 
identified particle is decomposed into components with respect 
to the measured reaction plane. In order to perform systematic studies 
the $p_{x}$ versus $p_{y}$ distributions are evaluated for different 
centrality and rapidity bins and for all identified particle species.
  The second method is based on the Fourier 
decomposition\cite{WA80-GUTB-90-B,SIM-VOLO-94-A}. The
azimuthal distributions of identified particles with respect to 
the reaction plane for all events are constructed. These distributions
are fitted with the function:
\begin{displaymath}
\frac{1}{N}\frac{dN}{d\Delta\Phi} = 1 + 
                                    2 v_1 \cos(\Delta\Phi) + 
                                    2 v_2 \cos(2\Delta\Phi)
\end{displaymath}
Where $\Delta\Phi$ is the azimuthal angle of the single particle 
with respect to the reaction plane. The strength of the collective
flow is then given by $v_1$ ($v_2$) for the directed (elliptic) flow.

  The flow measurements with respect to the reaction plane assume a 
perfect event plane determination, i.e. that the reaction plane 
angle could be obtained from the data exactly. In reality, the limited
detector resolution and effects like the finite 
number of detected particles produces a limited resolution in the 
measurement of the reaction plane angle. All observables that refer to 
the reaction plane must be corrected up to what they would be relative 
to the true event plane\cite{SIM-OLLI-97-A}. This correction is done 
by dividing the observable by the event plane 
resolution\cite{SIM-DANI-85-A,NA49-POSK-98-A}.

  Another effect which has to be taken into account, 
is the auto correlation effect. 
Naturally, there is a correlation between the azimuthal angle of a particle
with respect to the reaction plane, if this particle is included in 
the evaluation of the reaction plane angle. This auto correlation
is avoided by calculating for each particle the event plane angle of the 
remaining particles.

\section{Results}
\begin{figure}[htb]
\vspace*{-5mm}
\begin{minipage}[t]{77mm}
\psfig{figure=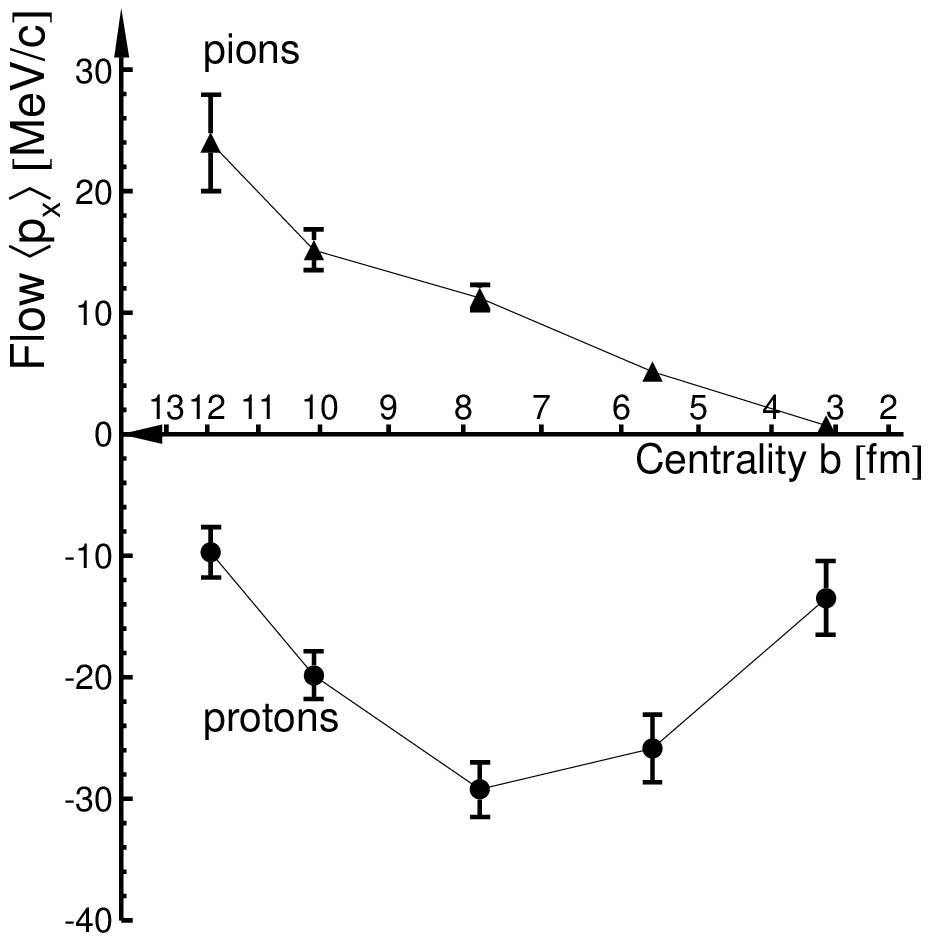,width=77mm}
\vspace*{-10mm}
\caption{\label{apxvswn} The average transverse momentum for protons
(circles) and pions (triangles) as function of centrality 
in terms of the impact parameter $b$.}
\end{minipage}
\hspace{\fill}
\begin{minipage}[t]{77mm}
\psfig{figure=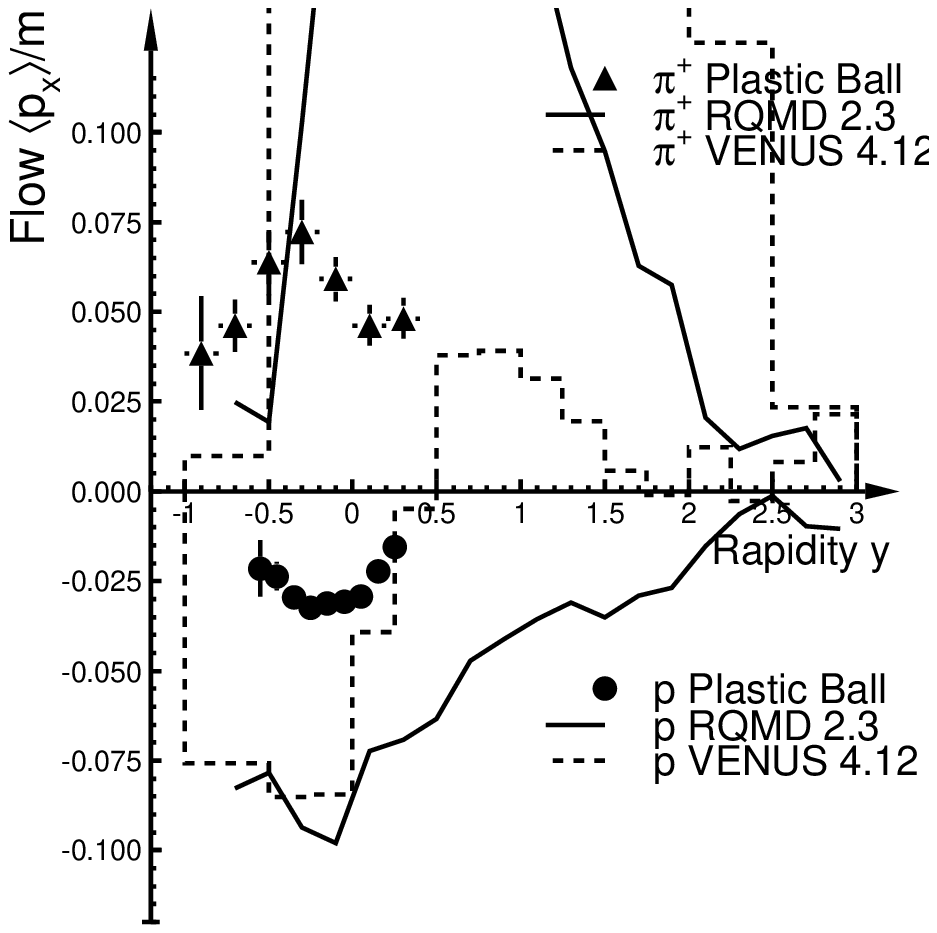,width=77mm}
\vspace*{-10mm}
\caption{\label{apxvsy} The average transverse momentum for protons
(circles) and pions (triangles) as function of rapidity $y$
compared to model predictions.}
\end{minipage}
\vspace*{-5mm}
\end{figure}

A systematic study of the dependence of the flow signal on centrality for 
protons and pions in terms of $\langle p_{x} \rangle$ is displayed
in figure \ref{apxvswn}.
The proton absolute momentum transfer in the reaction plane 
increases with the number of participants to a maximum 
$|\langle p_{x} \rangle|$ for semi-central collisions at an 
impact parameter of 
$b \approx 8\,\mbox{fm}$\cite{WA80-AGGA-98-B,QM99-PEIT-99-A} which is 
twice as large as that found for Au + Au collisions at AGS 
energies\cite{MISC-REIS-97-}.
For more head-on collisions the $|\langle p_{x} \rangle|$ decreases
again. In the limit of impact parameter zero the sideward flow vanishes 
due to symmetry.

In addition to the small flow effect of pions due to the thermal 
motion, pions are subject to absorption and rescattering
mainly through the delta resonance.
Thus they should show flow effects comparable to that
of protons\cite{SIM-BASS-95-A,SIM-LIBA-94-A}. Since the observed
$\langle p_{x} \rangle$ of pions is positive it indicates that 
the pions are preferentially emitted away from the target spectators.
This leads to the interpretation of an absorption of the pions in 
the target remnant, which appears as preferred emission toward
the other side.
If the apparent anti-flow is due to absorption, central collisions 
with little or no spectator matter should show no flow 
effect\cite{SIM-FILI-96-A} which is indeed seen for central events, where
the pion flow signal is compatible with zero. The effect in semi-central 
collisions is weak but grows with the impact parameter nearly linearly. 

 Figure \ref{apxvsy} shows the rapidity dependence of the average
transverse momentum for protons and pions in comparison with model
predictions. For protons a clear maximum flow at target rapidity
is evident in the data as well as in the simulations, though the
absolute height of the data cannot be reproduced by RQMD\cite{SIM-SORG-95-A}
or VENUS\cite{SIM-WERN-93-A}. 
There is also a large discrepancy between the pion data and the model
predictions, though the absolute height is approximately reproduced 
in the most backward rapidity regions.

\begin{figure}[htb]
\vspace*{-5mm}
\begin{minipage}[t]{77mm}
\psfig{figure=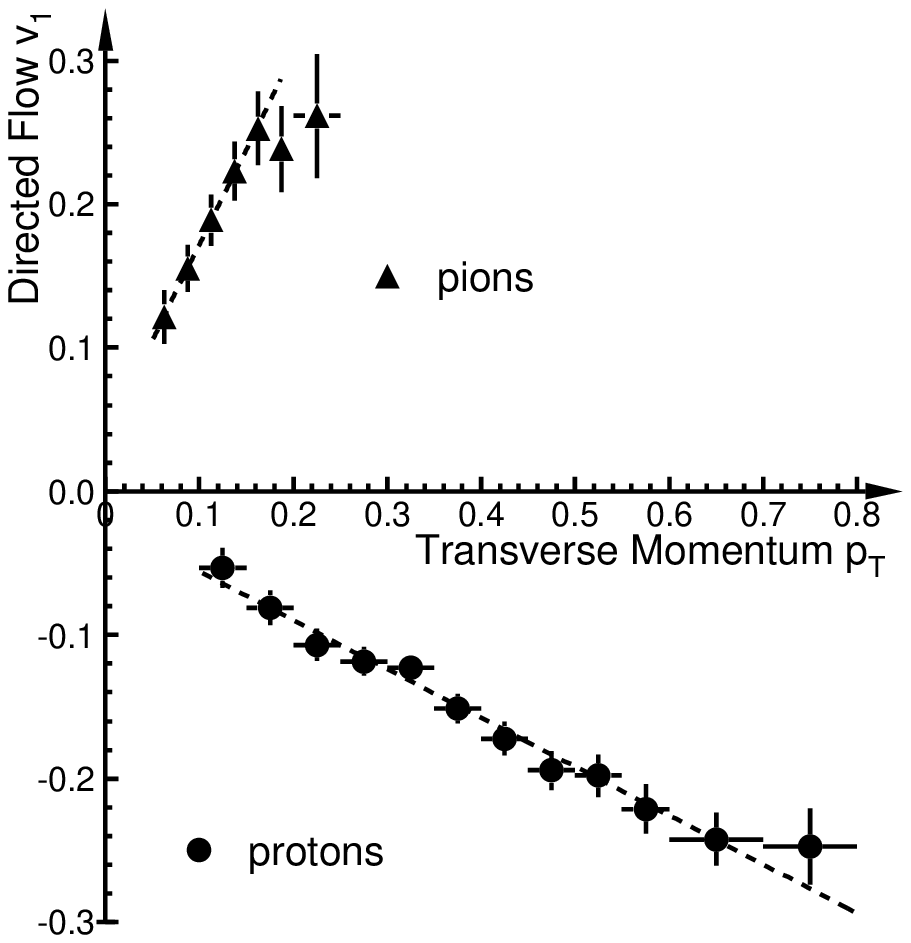,width=77mm}
\vspace*{-10mm}
\caption{\label{v1vspt} The directed flow in terms of $v_1$ for protons
(circles) and pions (triangles) as function of transverse 
momentum $p_T$.}
\end{minipage}
\hspace{\fill}
\begin{minipage}[t]{77mm}
\psfig{figure=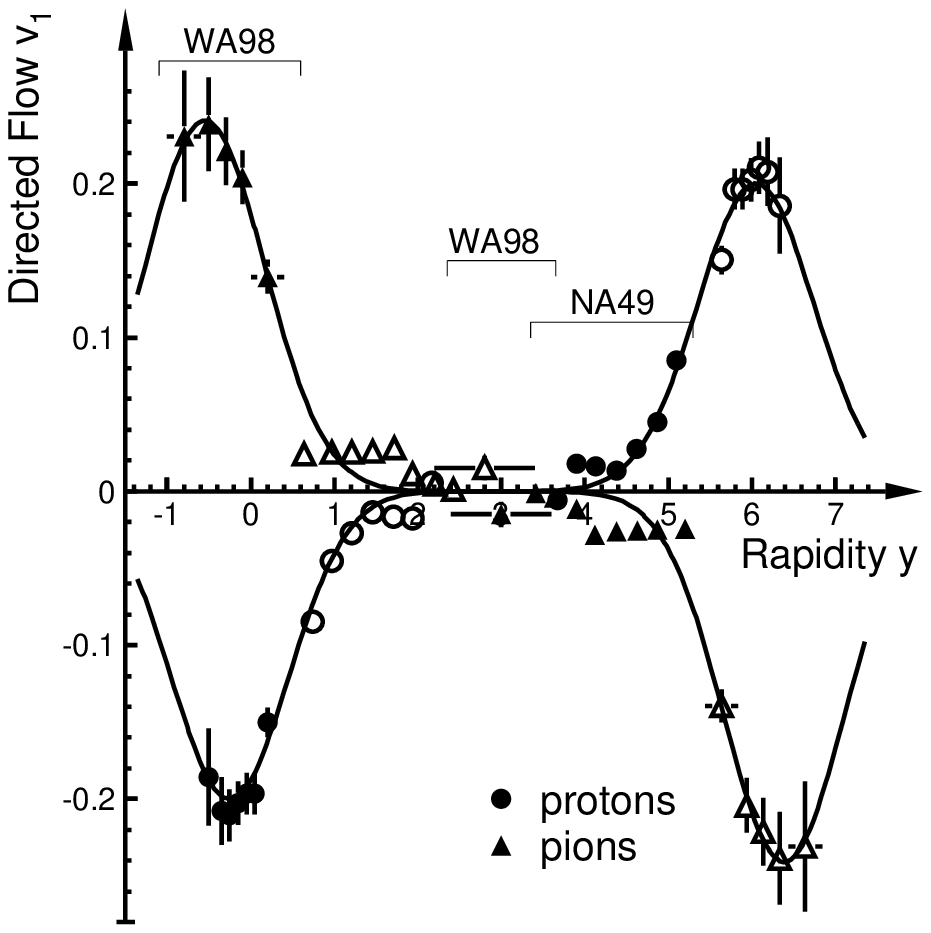,width=77mm}
\vspace*{-10mm}
\caption{\label{v1vsy} The flow parameter $v_1$ for protons
(circles) and pions (triangles) as function of rapidity $y$ 
in semi central collisions.}
\end{minipage}
\vspace*{-5mm}
\end{figure}

  The Fourier decomposition method also provides the transverse
momentum dependence of the flow strength. Over a wide range of 
$p_T$ the directed flow in terms of $v_1$ is well described by a 
linear function of the transverse momentum as depicted in figure
\ref{v1vspt}. The second harmonic or elliptic flow in terms of
$v_2$ is consistent with zero in the target rapidity range, hence
it is not plotted in the figure.

  Figure \ref{v1vsy} shows the rapidity dependence of the directed 
flow in terms of $v_1 = \langle \cos(\Delta\Phi) \rangle$. The filled
symbols represent measured data, while the open symbols are the data
reflected around midrapidity $y = 2.9$. Shown are proton (circles)
and pion (triangles) data from the Plastic Ball in the target
rapidity region. In addition, pion data measured with the tracking arm
in the WA98 experiment\cite{QM99-NISH-99-A} at midrapidity and data
near midrapitiy measured by the NA49 collaboration\cite{NA49-APPE-97-A}
are shown. It can be noticed that the directed flow of protons
as well as of pions has a maximum in the fragmentation regions. 

  The Plastic Ball data in the region $y < 0.5$ are fitted with Gaussian 
distributions. These Gaussian distributions, shown as solid lines, 
are reflected 
around midrapidity like the data and describe the data rather well. 
It should be emphasized that the midrapidity data
were not included in the Gaussian fit.
  The shape of the distribution appears different than that 
observed in heavy-ion
collisions at lower beam momenta\cite{WA80-DOSS-87-A,E877-BARR-96-A}, where 
the flow strength increases from zero at midrapidity linearly to the peaks at 
target and projectile rapidity. In 158 AGeV Pb + Pb collisions 
however, the peaks are Gaussian and only the tails extend to midrapidity.
It is conceivable that the S-shape curve obtained at lower beam energies 
could also be obtained by
a combination of two Gaussian distributions. In this case
the ratio of the relative width and the gap between the Gaussian peaks 
would be smaller so that a linear behaviour is found at midrapidity.
This suspicion was confirmed by a good agreement with a Gaussian
fit for the proton data
from 200 AMeV Au + Au collisions provided by the Plastic
Ball collaboration\cite{WA80-DOSS-87-A}.

  Hence for a complete description of the rapidity distribution of the 
collective flow $F$ the formerly used slope at midrapidity ($dF/dy|_{y=0}$)
is not sufficient. It is more reasonable to use the three parameters of 
the Gaussian distribution to describe the data. The peak position reflects 
the beam momentum, the peak height gives the strength of the flow and the 
width of the distribution provides information on how much the participants 
and the spectators are involved in the collectivity.

\end{document}